**Robust FCC solute diffusion predictions from ab-initio machine learning methods**

Henry Wu[1], Aren Lorenson[1], Ben Anderson[1], Liam Witteman[1], Haotian Wu[1], Bryce Meredig[2], Dane Morgan[1]

1. Department of Materials Science and Engineering, University of Wisconsin-Madison, Madison, USA
2. Citrine Informatics, Redwood City, CA 94061, USA

Abstract

We evaluate the performance of four machine learning methods for modeling and predicting FCC solute diffusion barriers. More than 200 FCC solute diffusion barriers from previous density functional theory (DFT) calculations served as our dataset to train four machine learning methods: linear regression (LR), decision tree (DT), Gaussian kernel ridge regression (GKRR), and artificial neural network (ANN). We separately optimize key physical descriptors favored by each method to model diffusion barriers. We also assess the ability of each method to extrapolate when faced with new hosts with limited known data. GKRR and ANN were found to perform the best, showing 0.15 eV cross-validation errors and predicting impurity diffusion in new hosts to within 0.2 eV when given only 5 data points from the host. We demonstrate the success of a combined DFT + data mining approach towards solving materials science challenges and predict the diffusion barrier of all available impurities across all FCC hosts.

1: Introduction

Atomic migration in solids governs the kinetics of many materials processes, including precipitation, high-temperature creep, phase transformation, and solution homogenization. A particular class of atomic diffusion is dilute impurity diffusion, which refers to the diffusion of a dilute solute in a host. Such diffusion is relevant in many materials applications as dilute solutes are common due to either undesired impurities or intentional dopants in materials. Due to its importance to materials science, large experimental catalogues of impurity diffusion measurements have been collected[1, 2], and more recently, first-principles predictions of dilute impurity diffusion coefficients have been conducted[3-8]. However, both experimental and theoretical approaches are limited by several drawbacks. Experimental diffusivities often vary significantly due to uncertainties introduced by different measurement techniques and other impurity effects in sample materials. In addition, experiments require significant diffusion kinetics to achieve good data, which generally limits them to be relatively high temperature (e.g., above about 50% of the melting temperature of the host)[9]. Finally, experiments are time consuming and expensive relative to first-principles calculations, as they require significant equipment and human interaction. First-principles calculations are an increasingly powerful tool for predicting dilute impurity diffusion, and compared to experiments can be done at a tiny fraction of the cost of equipment and human time. Furthermore, first-principles predicted energies are expected to be most accurate at lower temperatures, where vibrational and electronic excitations play a minor role, which suggests that these methods may have their best accuracy in temperature domains complimentary to experiments. First-principles methods bring with them significant approximations in both the fundamental energetics and in treating certain materials aspects, e.g., magnetic ordering, but agreement with experiments in large comparisons is generally quite good[8]. However, even with the speed of modern computers, a dilute diffusion coefficient calculation can take tens of thousands of CPU hours, which makes it expensive and time consuming to generate large databases. In particular, it would be beneficial to enable researchers to explore trends through searches of many thousands of possible solute-host combinations. In this work we use informatics approaches to develop more rapid ways to predict dilute diffusion coefficients in hosts.

We employ multiple machine learning (ML) algorithms to predict diffusion energy barriers for impurities in face-centered cubic (FCC) host systems. Our ML models train on more than 200 host-impurity diffusion systems calculated by Wu et al.[8] using first principles density functional theory (DFT). It has been long observed that strong correlations exist between impurity diffusion and elemental properties of the impurity and host, including melting temperature[10], cohesive energy[11], elastic constants[12], and atomic size[13]. ML methods are able to remove human bias and can decipher complex coupled property effects. ML have proven successful in many other materials problems, such as prediction of irradiation hardening in steels[14], melting temperature predictions of single and binary component solids[15], and the ground and excited state frontier orbital eigenvalues of molecules[16]. For diffusion in particular, Zeng et al.[17] have used linear regression and neural networks to fit experimental diffusivities. Our work will train additional model types on a comprehensive dataset from a single source, removing uncertainties associated with collecting a large experimental database from many sources. Our approach will also identify the physical qualities important to diffusion via analysis and optimization of a large array of model parameters.

In this paper, we will first detail the diffusion data used as well as the full set of training parameters (descriptors). We then briefly describe the four algorithms we consider, which are linear regression, decision tree, Gaussian kernel ridge regression, and artificial neural network. We use standard statistical analysis methods to optimize descriptor sets for each method and determine cross-validation errors and predictive capabilities for new/unknown host systems.

2: Methods
2.1: Dataset

The dataset we use for training and testing contains 218 host-impurity diffusion activation energy barriers calculated by DFT methods in Face-Centered Cubic (FCC) metallic hosts[8]. This include 30-40 impurity diffusion barriers for each of 5 major hosts, Al, Cu, Ni, Pd, and Pt; and less 10 impurities for each of 4 minor hosts, Au, Ca, Ir, and Pb. Each diffusion activation barrier for impurity, $i$, diffusing in host, $h$, has been normalized by subtracting off the host self-diffusivity, $E_{h-h,0}$, by the following equation,

$E_{h-i} = E_{h-i,0} - E_{h-h,0}$ Eq.(1)

where, $E_{h\text{-}i,0}$ is the unnormalized activation barrier and $E_{h\text{-}i}$ is the normalized activation barrier. This normalization is chosen based on the observation that the difference between the impurity diffusion activation barrier within a host and the host's self-diffusivity is captured especially well by DFT. We wish to capture this agreement with our models in order to best predict impurity diffusivities within new host elements. In general, the host self-diffusivities are well known from experimental measurements[1, 2], so Eq.(1) can be used with a ML prediction to get the true impurity diffusion activation energy.

To illustrate the value of using barriers relative to hosts we show in Figure 1 our DFT dataset of normalized activation barriers plotted against normalized molar volumes. We can see a clear trend for decreasing normalized activation barriers for more positive normalized molar volumes. This correlation would not be clear if the impurity diffusion barriers were plotted directly as calculated, without referencing to the host. This trend is also observed in the literature that impurity atoms larger than the host atom diffuse faster than expected[4, 18, 19] and in many cases faster than the host self-diffusivity. This effect can result from both an increased binding between the large impurity atom and vacancies in the host due to misfit strain, as well as a decrease in the impurity-vacancy migration barrier. An attractive binding energy manifests as a reduction in the vacancy formation energy while a decrease in the impurity-vacancy migration barrier allows the vacancy to move faster, both leading to a reduced overall impurity diffusion activation energy.

We characterize each host-impurity system with a feature vector of 111 descriptors, which include elemental properties for the host and impurity atoms, as well as difference or ratio of properties between host and impurity atoms. These descriptors were compiled mainly from the database included in Wolverton Research Group's Magpie[20] with a few additional descriptors from literature[21-24] and web-based elemental databases[25]. We removed 24 constant descriptors (i.e. constant value for all host-impurity pairs) and those with strong linear correlations (greater than 0.98) with another descriptor from our descriptor set. When two descriptors were correlated strongly we tried to keep the most physically intuitive descriptor. The remaining 111 descriptors are normalized linearly to between 0 and 1 corresponding to their minimum and maximum values, respectively, and form the initial feature vector for four different ML algorithms. The complete database of all activation energies and all descriptors for all the host-impurity pairs is given in the Supplemental Information.

2.2: Machine learning methods

The four machine learning algorithms we will utilize are: linear regression (LR), decision tree (DT), Gaussian kernel ridge regression (GKRR), and artificial neural network (ANN). For LR, DT, and GKRR, we have used the implementation provided by scikit-learn[26], a python code package of machine learning tools. For ANN, we have adopted a python implementation from Stephen Welch[27]. All variable names below follow those used in the codes we have used in the modeling and their definitions can be found in the respective code documentations for LR[28], DT[29], GKRR[30], ANN[27].

LR is a common and simple model for regression analysis. An LR model assumes the relationship between the dependent variable, diffusion activation barrier, and the independent variables, feature vector of descriptors, is linear. This assumption gives LR models their simplicity and leads to easily understandable analysis and rapid fitting.

However this simplicity is also the model's drawback since non-linear effect cannot be captured, and the model can be very sensitive to outliers. We include LR as a baseline test for the other ML models.

The DT method learns about and predicts the data by producing a flowchart with decision questions at each node, which can branch out into additional decision nodes, terminating at an outcome. We limit the decisions to binary, numerical separations on the feature descriptors, while the outcomes are discrete binned values for the diffusion activation barriers. We have chosen to optimize two decision tree model parameters, max_depth, the maximum depth of a tree, and min_sample_split, the minimum number of samples required to split a node with a decision. Both of these parameters put limits how many decision questions can be asked as well as how finely divided the outcome numerical bins can be. Decision trees can overfit the data if either max_depth is too large or if min_sample_split is too small. We use cross-validation RMSE to optimized these parameters and find: max_depth = 5 and min_samples_split = 8.

GKRR fits local linear functions to data characterized by their Gaussian kernel distance. This method allows for efficient data fitting to highly non-linear functions. To use GKRR we set kernel='rbf' and optimize two model parameters, alpha and gamma. Alpha is the regularization parameter and controls how much local function coefficients are penalized; a small alpha will give a better fit to the given local data but can lead to overfitting. Gamma acts as an inverse length scale in the Gaussian kernel and controls the local range over which the functions can change. A smaller gamma will give a more average fit over a larger range of data while a larger gamma will give a fit that is more dependent on the closest data points. We use cross-validation RMSE to optimized these parameters and find: alpha = 0.031623 and gamma = 0.14678.

ANN models data with a network of interconnected layers of artificial neurons. The feature descriptors each feed into a neuron in the input layer, which are then propagated with fitted weights into various neurons in a hidden layers. The ANN may consist of multiple hidden neuron layers before terminating in the output neuron layer. The output of each neuron propagates forward through an activation function of their weighted inputs. We have chosen to use only a single hidden layer, which consists of half of the number of neurons in the input layer, changing the number of neurons in both layers as the number of descriptors change. In addition, to reduce in the stochastic nature of the ANN method, we train 50 networks and take an ensemble average of the top 5 performing networks. The network performance is determined by its root-mean-squared error (RMSE) on validation data contained wholly within the training data and separate from any cross-validation testing data.

2.3: Criteria for optimization

To judge the performance of different models we use leave-out-20% cross-validation (LO20% CV). In LO20% CV our full dataset is randomly partitioned in two, 80% is used to train a model and 20% is left out to test the resulting model. For only the ANN method, 20% of the training data is further partitioned to be validation for the ensemble averaging. The root-mean-square error (RMSE) between the actual values in the test set and model predictions are then calculated. This procedure is repeated 100 times (20 times for ANN due to the longer runtime) with different random partitions and the RMSE from each of the CV tests is averaged together and reported in this text and plotted in all figures. This average will simply be referred to as the CV RMSE.

While we start with 111 descriptors for our data, ideally we only want to work with descriptors that are important for predicting diffusion, which may be different for each ML method. The goal for optimizing the descriptor set is to reduce the time required for fitting and to produce a relatively short list of descriptors to aid in human analysis of which physical factors are most important. Descriptor reduction can also be helpful to filter out similarly behaving descriptors that are essentially redundant to improve model performance. We accomplish the reduction for each of the four ML methods separately in a three-step procedure. Step one generates an ordered list of all descriptors based on their initial importance to the model's LO20% RMSE. For LR and GKRR this is done using forward selection of descriptors; where a single descriptor is selected based on how much it reduces the model RMSE with subsequent descriptors being selected based on the same criteria when pooled together with all previous descriptors. For DT, the sum of each descriptor's Gini impurity index is sorted to produce the ordered list. For ANN, the ordered list is sorted from the summed absolute weights propagating from each descriptor's input neuron to the output neuron through all hidden layer neurons. Step two reorders the initial ordered descriptor list according to which descriptor incrementally reduced the CV RMSE the most when added to the previous pool of descriptors. Finally, step three selects a cutoff for the optimal set of descriptors for each method at the point where the reduction in the average CV RMSE plateaus or when the reduction becomes indistinguishable from noise, taken to be the standard deviation of the CV RMSE.

3: Results and Discussion
3.1: Descriptor optimization (dimensionality reduction)

Figure 2 shows the CV RMSE curves for each ML method plotted against the ordered descriptor list for each particular method. The descriptor order for each method has been determined from the procedure detailed in Sec. 2.3. Each point is an average over 100 separate LO20% CV runs (20 separate runs for ANN), the standard deviation for those runs is plotted as the shaded area around each respective curve. We can see that, in general, this ordered list of descriptors improves the performance of each model monotonically with increasing descriptors. This monotonic trend shows that the re-ordering procedure has done a good job of removing harmful or unnecessary descriptors.

We can also make several observations regarding each ML method from Figure 2. LR and DT performed much worse than GKRR and ANN and required 8 and 5 descriptors to do so, respectively. This poor performance was expected from these very simple methods but does serve as baselines for the other methods, showing to what extent more complex nonlinear models can improve the results. DT shows the largest standard deviation out of all the methods, which indicates higher sensitivity to training sets and overfitting. The curve for GKRR appears similar in shape to the curve for LR, a smoothly decreasing RMSE with increasing descriptors, and reaches optimal value at 23 descriptors. The similarity between GKRR and LR can be further seen as they both selected the melting temperature of the impurity as the most important descriptor. Training to only this first descriptor, both GKRR and LR give almost the same CV RMSE. However, due to its more flexible non-linear nature, GKRR significantly outperforms LR with further descriptors. ANN starts off poorly for the first few descriptors but quickly improves to be comparable with GKRR at 14 descriptors. The observed noise in the ANN curve for smaller numbers of descriptors is due to both the fewer number of CV runs and the stochastic nature of neural networks.

Table I shows the optimal list of descriptors for each of the four ML methods. The exact description for each descriptor can be found in the Supplemental section. Several interesting physical observations can be seen from Table I. There are many expected descriptors in the list, e.g., those that have to do with atomic size (atomic radii, lattice constant, and atomic volume) and bond strength (ground-state energy, melting temperature, and elastic constant). Some descriptors that were not expected include ones describing relationships within the periodic table (periodic column, periodic row) and electronic states (number of valence electrons, oxidation states). GKRR shares 4 descriptors with LR and 6 descriptors with ANN, and these three methods mutually share 3 descriptors: melting temperature of the impurity, number of filled d valence orbitals for the host, and the ground state DFT energy difference between the impurity and host. We note that the impurity's melting temperature was the number one descriptor for all three of these ML methods, and absent from DT. This suggests that our procedure may not be sufficient to optimize the descriptor set for DT. In addition, the given list of descriptors should not be taken to be unique as there are many correlated descriptors that may perform equally well. However, the physical property categories that correspond to the identified descriptors, e.g., size and bond strength, are expected to play a dominant role in any effective descriptor list.

3.2: Full fit

Figure 3 shows the fitted impurity activation barriers from each of the four ML methods compared against the DFT activation barriers. These fitted results come from using the entire DFT dataset as the training set for each ML method. The fitted linear equation for LR and fitted tree for DT can be found in the supplemental information. The fitted GKRR model can be replicated from the provided alpha and gamma parameters. For ANN, the fitting results from an ensemble of the top 5 out of 50 networks ordered by separating out random 20% of the full dataset as internal validation sets; the full network of weights will not be provided, as it is unwieldy to use.

Figure 3 shows clear bias from LR: This model underestimates very high and very low impurity barriers, while overestimating those barriers in between. Several horizontal clusters are readily apparent in the DT plot; these clusters are a direct consequence which result from the numerical binning required for decision tree regression. We also see DT predictions becoming increasingly worse at higher diffusion barriers, predicting the same numerical barrier for wider and wider ranges of actual barriers. No bias can be seen for GKRR and ANN.

Table II shows a comparison between the optimal LO20% CV RMSE from Figure 2 and the fitting RMSE from Figure 3. In general, the cross-validation error should always be higher than the fitted error, and this is this case here for all four methods. The CV RMSE should approach the fitting error in an ideal case, while much higher CV RMSE may be indicative of overfitting. For LR we see that these two errors are comparable, with the fitting error being only 4% lower than the CV RMSE. However,

DT, GKRR, and ANN all show significant changes, with the fitting error being approximately 28%, 32%, and 38% lower than the CV RMSE, respectively. While further work on these methods may improve CV RMSE and avoid overfitting, such detailed optimization is outside the scope of this work.

We have utilized all four, fully trained, ML methods to predict the diffusion barrier of all impurities in all FCC host elements. The full prediction dataset is included in the supplemental information and accessed on our interactive web app[31]. Certain host-impurity combinations have not been predicted because not all atomic properties are known for every element in the periodic table, and because each of the ML methods utilize different atomic features. We stress that these predictions should not be directly used without serious consideration of many issues. For example, we would expect many of the light elements to diffuse interstitially, rather than the vacancy-mediated model that is at the base of our DFT diffusion calculations. More trust should be assigned to predictions for systems that are most similar to the training data, namely when both host and impurity are part of the transition metals.

3.3: Extrapolation to new hosts

While CV provides a useful overall assessment of ML model accuracy, when considering the ability to extrapolate to new host-impurity pairs it is likely that entirely new hosts will be particularly challenging. Therefore, we also test the predictive ability of each model for new hosts as a function of how much data for that host is included in the fit. Figure 4 shows results for predicting impurity activation barrier in new/unknown host systems not included or only partially included in the fitting for all four methods. We will refer to these prediction errors as the leave-host-out RMSE. We train the ML models with data from all but one of the major host systems (Al, Cu, Ni, Pd, or Pt) and predict the impurity diffusion barriers of the left-out host. We also add increasing number of random impurities from the left-out host into the training data to see how the leave-host-out RMSE changes with additional information about the unknown host. This test is meant to tell us the expected error from each method when predicting diffusion values for new host systems where we might have no data or limited data. We see from Figure 4 that both LR and DT give relatively flat curves even with increasing number of impurities from the left-out host in the training data. This insensitivity is expected from these two methods since both models have relatively few parameters compared to GKRR and NN, and therefore fit a highly averaged model over the bulk of the available data and thus will not change significantly with small additions of new data. On the other hand, both GKRR and ANN appear to improve their unknown host predictions with additional information about the host. This effect is most apparent in Al where both GKRR and ANN give errors larger than 0.6 eV when they include no training data from the Al host. This improves substantially when more and more impurities from Al are added into their training data, down to 0.25 eV when 12 Al impurities have been included. The other major hosts, Cu, Ni, Pd, and Pt all show prediction errors <0.2 eV when including fewer than 5 additional impurities. We believe Al is the most challenging case because it is the smallest host element and does not contain d-shell electrons, both of which affects multiple descriptors used for GKRR and ANN (e.g. atomic radii, d-electron valence, etc.). This suggests that for other FCC hosts, we expect only Ca, also small and lacking d-electrons, would require large number of additional calculated impurity diffusion barriers to reach good prediction accuracy.

4: Conclusion

      We have trained four separate machine-learning algorithms for predicting impurity diffusion activation barriers in FCC hosts and have made diffusion barrier predictions for all impurities across all FCC hosts. Over the course of model training, we have identified a set of impurity and host properties that are most important for determining the diffusion activation energy (relative to the host). These properties include ones know to be important for diffusion, such as atomic size, melting temperature, atomic energy, etc.; as well as unexpected ones such as the number of valence electrons and periodic row/column numbers. The descriptors that appear most frequently included the melting temperature of the impurity, the host and impurity DFT ground state energies, and the number of filled d valence orbitals for the host. While the frequent appearance suggests that these descriptors are particularly important, there are many similar descriptors in our lists and it is difficult to say definitely which are most effective for describing diffusion. Additional studies could explore determining the best classes, or unique small sets, of descriptors, but this is beyond the scope of the present work.

      We find that linear regression and decision tree, as very simple high-bias, low-variance methods, perform worst in cross-validation, but are otherwise quite stable with only a few descriptors. While there is more room for decision trees to improve with sophisticated tree fitting methods or ensembles of trees (e.g. gradient boosted, random forest), they were not explored in this work. Our best methods, Gaussian kernel ridge regression (GKRR) and artificial neural network (ANN) both give low cross-validation errors around 0.15 eV, approaching DFT and experimental accuracies. When predicting new host systems, both GKRR and ANN give an average of 0.2 eV errors after including just 5 impurities from the new host system in the training data. Impurity diffusion predictions for Al host gives much higher errors, >0.25 eV even with 10 impurities in the training set. This suggests that more work is needed to optimize GKRR and ANN as well as their descriptors to enable low error extrapolation into all unknown hosts with accuracy comparable to DFT. However, if one can tolerate errors between 0.2-0.3 eV, then the present GKRR and ANN models can robustly predict diffusion activation barriers across FCC hosts for many impurities. These predictions must be used with caution for a few reasons. First, the impurities must diffuse by a vacancy mechanism (as this is all that is modeled in this database here). Furthermore, the impurities were generally chosen to be common metal impurities or to form connected regions on the periodic table to allow searches for trends, and have not been chosen to assure all impurity chemical aspects are well represented. Therefore, some impurities with dramatically different chemistry than those considered here might be predicted quite poorly, e.g., He impurity diffusion. This work illustrates some of the strengths and weaknesses of major ML methods on a realistic materials dataset of kinetic properties. While the models are potentially accurate enough for extrapolating qualitative trends, they also show clear limitations in extrapolating to new hosts and uncertainty in extrapolation to novel impurity chemistries. Developing models that are more robust for all relevant systems is an important task for diffusion data and materials data in general.

**Acknowledgements**
We acknowledge that author Henry Wu was funded by the NSF Software Infrastructure for Sustained Innovation (SI2) award No. 1148011.

**References**

[1] Smithell's Metals Reference Book, 7th ed., Butterworth-Heinemann, Oxford, 1998.
[2] Pergamon Materials Series: Self-diffusion and Impurity Diffusion in Pure Metals, Pergamon2008.
[3] N. Sandberg, R. Holmestad, First-principles calculations of impurity diffusion activation energies in Al, Physical Review B 73(1) (2006) 014108.
[4] D. Simonovic, M.H.F. Sluiter, Impurity diffusion activation energies in Al from first principles, Physical Review B 79(5) (2009) 054304.
[5] M. Mantina, Y. Wang, L.Q. Chen, Z.K. Liu, C. Wolverton, First principles impurity diffusion coefficients, Acta Materialia 57(14) (2009) 4102-4108.
[6] S. Huang, D.L. Worthington, M. Asta, V. Ozolins, G. Ghosh, P.K. Liaw, Calculation of impurity diffusivities in α-Fe using first-principles methods, Acta Materialia 58(6) (2010) 1982-1993.
[7] S. Ganeshan, L.G. Hector Jr, Z.K. Liu, First-principles calculations of impurity diffusion coefficients in dilute Mg alloys using the 8-frequency model, Acta Materialia 59(8) (2011) 3214-3228.
[8] H. Wu, T. Mayeshiba, D. Morgan, High-throughput ab-initio dilute solute diffusion database, Scientific Data 3 (2016) 160054.
[9] H. Mehrer, Diffusion in Solids, Springer-Verlag Berlin Heidelberg2007.
[10] G. Neumann, W. Hirschwald, Impurity Diffusion in F.C.C. Metals, physica status solidi (b) 55(1) (1973) 99-111.
[11] D. Turnbull, R.E. Hoffman, A correlation of data on diffusion of solutes in face-centered cubic metals, Acta Metallurgica 7(6) (1959) 407-410.
[12] A. Ferro, Theory of Diffusion Constants in Interstitial Solid Solutions of b.c.c. Metals, Journal of Applied Physics 28(8) (1957) 895-900.
[13] G.M. Hood, An atom size effect in tracer diffusion, Journal of Physics F: Metal Physics 8(8) (1978) 1677.
[14] G.W. Colin, C. Geoff, K. Richard, Prediction of yield stress in highly irradiated ferritic steels, Modelling and Simulation in Materials Science and Engineering 16(2) (2008) 025005.
[15] A. Seko, T. Maekawa, K. Tsuda, I. Tanaka, Machine learning with systematic density-functional theory calculations: Application to melting temperatures of single- and binary-component solids, Physical Review B 89(5) (2014) 054303.
[16] K. Hansen, F. Biegler, R. Ramakrishnan, W. Pronobis, O.A. von Lilienfeld, K.-R. Müller, A. Tkatchenko, Machine Learning Predictions of Molecular Properties: Accurate Many-Body Potentials and Nonlocality in Chemical Space, The Journal of Physical Chemistry Letters 6(12) (2015) 2326-2331.
[17] Y. Zeng, K. Bai, High-throughput prediction of activation energy for impurity diffusion in fcc metals of Group I and VIII, Journal of Alloys and Compounds 624 (2015) 201-209.
[18] I.L. Lomaev, D.L. Novikov, S.V. Okatov, Y.N. Gornostyrev, S.F. Burlatsky, First-principles study of 4d solute diffusion in nickel, Journal of Materials Science 49(11) (2014) 4038-4044.
[19] A. Janotti, M. Krčmar, C.L. Fu, R.C. Reed, Solute Diffusion in Metals: Larger Atoms Can Move Faster, Physical Review Letters 92(8) (2004) 085901.

[20] L. Ward, A. Agrawal, A. Choudhary, C. Wolverton, A general-purpose machine learning framework for predicting properties of inorganic materials, Npj Computational Materials 2 (2016) 16028.
[21] E. Clementi, D.L. Raimondi, W.P. Reinhardt, Atomic Screening Constants from SCF Functions. II. Atoms with 37 to 86 Electrons, The Journal of Chemical Physics 47(4) (1967) 1300-1307.
[22] J.C. Slater, Atomic Radii in Crystals, The Journal of Chemical Physics 41(10) (1964) 3199-3204.
[23] R.J. Gillespie, The Periodic-Table, Chemistry in Britain 22(5) (1986) 414-415.
[24] C. Kittel, Introduction to Solid State Physics,, 8 ed., John Wiley & Sons, Inc, Hoboken, NJ, 2005.
[25] Chemicool Periodic Table. <http://www.chemicool.com/ >, 2016).
[26] F. Pedregosa, G. Varoquaux, A. Gramfort, V. Michel, B. Thirion, O. Grisel, M. Blondel, P. Prettenhofer, R. Weiss, V. Dubourg, J. Vanderplas, A. Passos, D. Cournapeau, M. Brucher, M. Perrot, E. Duchesnay, Scikit-learn: Machine Learning in Python, J. Mach. Learn. Res. 12 (2011) 2825-2830.
[27] S.C. Welch, Neural Network Demystified. <https://github.com/stephencwelch/Neural-Networks-Demystified >, 2016).
[28] scikit-learn, Linear Regression. <http://scikit-learn.org/stable/modules/generated/sklearn.linear_model.LinearRegression.html >, 2016).
[29] scikit-learn, Decision Tree Regressor. <http://scikit-learn.org/stable/modules/generated/sklearn.tree.DecisionTreeRegressor.html >, 2016).
[30] scikit-learn, Kernel Ridge. <http://scikit-learn.org/stable/modules/generated/sklearn.kernel_ridge.KernelRidge.html >, 2016).
[31] H. Wu, T. Mayeshiba, E. Boyer, D. Morgan, Dilute Solute Diffusion Database Predicted from Machine Learning. <http://mldiffusiondata.materialshub.org >, 2017).

**Figures and Tables**

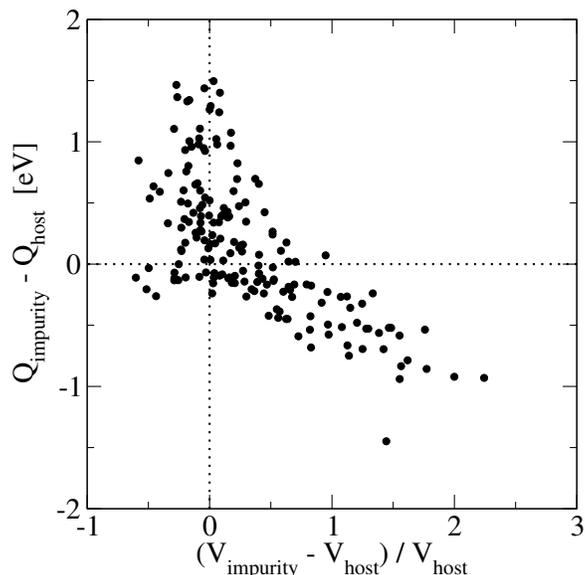

Figure 1. Trend of normalized impurity diffusion activation barriers plotted against the normalized impurity volume.

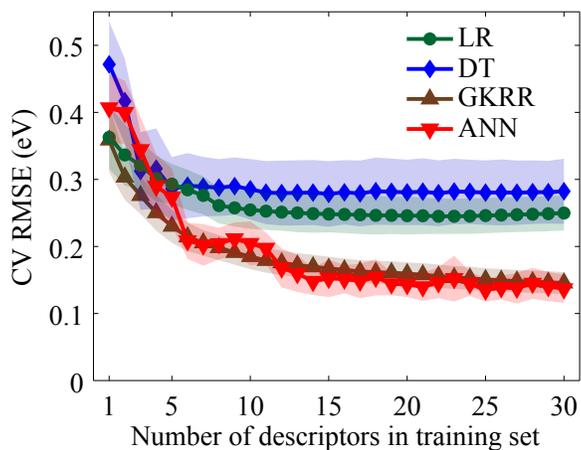

Figure 2. Change in the leave-out-20% cross-validation RMSE for each of the four machine learning methods with increasing number of descriptors included in the training set. The average RMSE over 100 cross-validation runs for each method (20 for ANN) are for each method are denoted by connected symbols surrounded by a shaded area representing one standard deviation error bars.

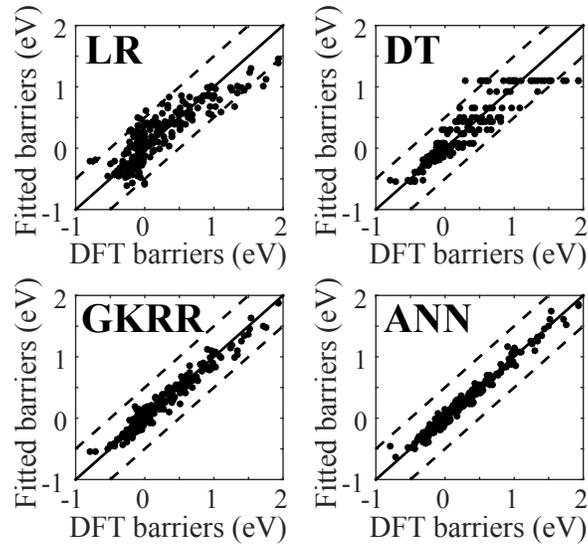

Figure 3. Fitted impurity diffusion activation barriers from each of the four machine learning methods plotted against the actual impurity diffusion activation barriers. The y=x line is shown as a solid line, the surrounding dashed lines represent ±0.5 eV shifts from the y=x line.

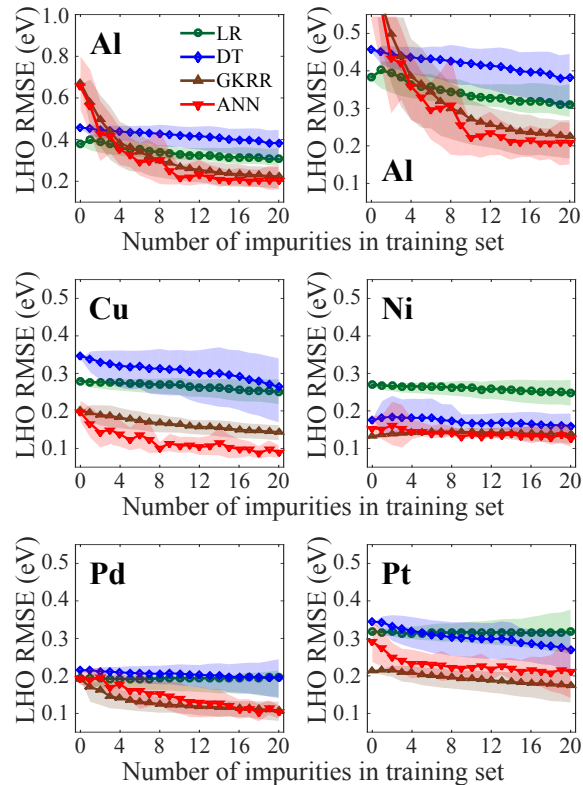

Figure 4. Change in the leave-host-out (LHO) RMSE for each of the four machine learning methods with increasing number of impurities from the left out host included in the training set. Results for each of the left out hosts, Al, Cu, Ni, Pd, and Pt, are plotted in separate plots. The upper right plot shows a zoomed in section of the Al plot. The average

RMSE over 100 runs for each method (20 for ANN) are denoted by connected symbols surrounded by a shaded area representing one standard deviation error bars.

Table I: The most important descriptors for each of the four machine-learning methods. Within each method, the descriptors are listed in ranked order of importance from left-to-right and then top-to-bottom. A full description of each descriptor named here is available in the Supplemental section.

| Method | Optimized descriptor set | | | | | |
|---|---|---|---|---|---|---|
| LR | MT (K) I | GSefflatcnt D | MT (k) R | NpUnfilled H | IsDB B | GSEnergy D |
|  | NpValence H | Electronegativity H |  |  |  |  |
| DT | GSefflatcnt D | SpaceGroup I | MinOx I | MinOx D | NdUnfilled D |  |
| GKRR | MT (K) I | NdValence H | MaxN H | NdUnfilled I | IsDB B | BM I |
|  | GSEnergy D | MaxOx I | NsValence H | AN D | GSEnergy I | MinOx H |
|  | BccMm D | AREmp (pm) H | NpValence D | MendeleevNumber I | Row H | Col D |
|  | IsMd I | BM D | ARCalc H | Col I | NdValence I |  |
| ANN | MT (K) I | NdValence H | IsDB I | GSVol D | GSEnergy I | AREmp (pm) I |
|  | BccVolDi H | NdUnfilled D | BM D | GSEnergy D | IsDB H | MaxN H |
|  | Col(M) I | CovRad H |  |  |  |  |

Table II: Comparison between the average leave-out-20% cross-validation RMSE and the fitting RMSE for each of the four machine learning methods evaluated with their optimal set of descriptors, respectively. Uncertainties represent one standard deviation over 100 cross-validation runs for each method (20 for ANN).

|  | LR | DT | GKRR | ANN |
|---|---|---|---|---|
| LO20% CV RMSE (eV) | 0.261±0.027 | 0.286±0.047 | 0.154±0.017 | 0.148±0.020 |
| Fitting RMSE (eV) | 0.251 | 0.207 | 0.105 | 0.092 |